\documentstyle[11pt,newpasp,twoside, epsf]{article}
\def\edcomment#1{\iffalse\marginpar{\raggedright\sl#1\/}\else\relax\fi}
\reversemarginpar

\begin{document}
\title{The Luminosity Function of 80 Abell Clusters from the CRoNaRio catalogues}
 \author{Piranomonte S., Longo G., Andreon S., Puddu E.}
\affil{Osservatorio Astronomico di Capodimonte, Via Moiariello 16, 80131 Napoli}
\author{Paolillo M.}
\affil{D.S.F.A., Universit\'a di Palermo, Via Archirafi 36, 90123 Palermo}

\author{Scaramella R.}
\affil{Osservatorio Astronomico di Roma, Via Frascati 33, 00040 Roma}

\author{Gal R., Djorgovski, S.G. }
\affil{Department of Astronomy, Caltech, USA}
\begin{abstract}
We present the composite luminosity function (hereafter LF) of galaxies for 80 Abell clusters studied in our survey of the Northern Hemisphere, using DPOSS data in the framework of the CRoNaRio collaboration. Our determination of the LF has been computed with very high accuracy thanks to the use of homogeneous data both for the clusters and the control fields and to a local estimate of the background, which takes into account the presence of large-scale structures and of foreground clusters and groups.
The global composite LF is quite flat down to $M^*+5$ and it has a slope 
$\alpha\sim-1.0\pm0.2$ with minor variations from blue to red filters, and $M^*\sim-21.9,-22.0,-22.3$ mag ($H_0=50$ km s$^{-1}$ Mpc$^{-1}$) in $g, r$ and $i$
filters, respectively (errors are detailed in the text).
We find a significant difference between rich and poor clusters supporting the existence of an LF's dependence on the environment.
\end{abstract}


\section{Introduction}
The galaxy luminosity function measures the relative frequency of galaxies as a function of luminosity.
Cluster LF's can be determined as the statistical excess of galaxies along the line of sight of the clusters with respect to control field directions due to the fact that clusters appear as overdensities with respect to the field. This approach assumes that the contribution of the fore-background galaxies along the line of sight of the clusters is equal to an ``average'' value. An hypotesis that is rather weak since very often nearby groups, clusters or superclusters happen to lay in same the direction and therefore affect the determination of the LF. 

This  problem is even more relevant when sampling the cluster outskirts due to: i) the low galaxy density of these regions is strongly affected even by small contaminations; ii) the large observing area which makes more probable the presence of contaminating structures. These outer regions are very relevant due to the fact that they are the putative places for galaxy evolution to occur (van Dokkum et al. 1998).

Non zero correlation function makes therefore very time consuming the accurate determination of LF's using traditional CCD imagers (due to their small field of view) and therefore leads most authors to use comparison average field counts taken from the literature. These average counts are usually obtained from regions of the sky completely unrelated to the cluster's position. Alternative routes are either to observe small comparison fields or to recognise individual cluster membership either on spectroscopic or morphological grounds.

Wide field (hereafter WF) imagers such as Schmidt plates or large format CCD's are the ideal tools to perform accurate determinations of LF's for statistically significant samples of clusters. In this paper we present results from a long term project aimed to derive LF's for a large sample of Abell clusters selected accordingly to the criteria described below.

The work is done in the framework of the CRoNaRio collaboration aimed to produce the first complete catalogue of all object visible on the Digitised Palomar Sky Survey (hereafter DPOSS).

\section{The CRoNaRio Project and the cluster sample}
The CRoNaRio Project is a joint enterprise among Caltech and the astronomical observatories of Roma, Napoli and Rio de Jainero, aimed to produce the first general catalogue of all objects visible on the DPOSS. The final Palomar-Norris North Sky Catalogue will include astrometric, photometric (in the three Gunn-Thuan bands $g$, $r$ and $i$) and rough morphological information for an estimated number of $2 \times 10^9$ stars and $5 \times 10^7$ galaxies (Djorgovsky et al.1999).

Our final goal is to derive individual cluster LF's and therefore a statistically robust cumulative LF for all Abell (1958) clusters fulfilling the following criteria: the cluster must fall in a plate triplet with available individual photometric zero points, it must not be close to the edges of a plate and it must have at least one reliable spectroscopic redshift estimate. Moreover clusters must not present anomalous structures such as double density peaks, discordant redshift determinations etc.

In what follows we discuss the results for a subsample of 80 clusters, an extension of Paolillo et al.(2000).

\section{Individual LF determination}
In order to compute the cluster LF the first step is to subtract the fore-background contamination. The method, together with the errors involved in the subtraction process, is detailed in Paolillo et al. (1999) and in Paolillo et al. (2000).

Field counts are measured around each cluster, thanks to the wide field coverage of DPOSS plates, after removing density peaks.

A ``local field'', measured all around the studied clusters, is the adopted estimate of the
background counts in the cluster direction.  It is a better measure of the contribution of
background galaxies to counts in the cluster direction than the usual ``average'' field,
since it allows to correct for the presence of possible underlying large-scale structures
both at the cluster distance and in front or behind it.


\begin{figure}[t]
\begin{center}
\epsfysize=8cm 
\epsfxsize=8cm 
\hspace{0.cm}\epsfbox{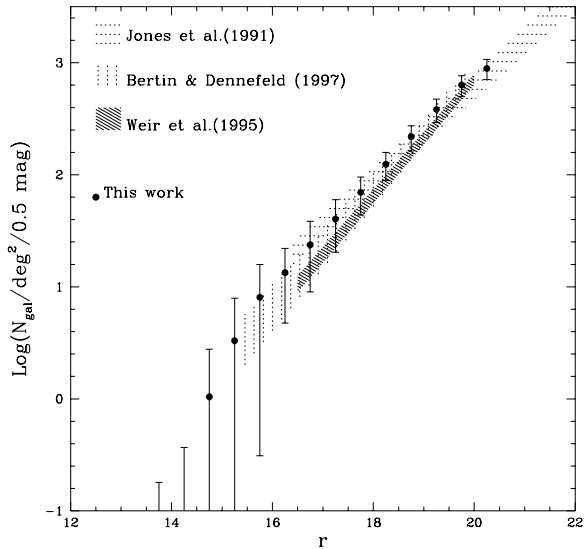} 
\end{center}
\caption[h]{\it{Comparison of our background counts with some 
previous determinations.}}
\end{figure}

In Figure 1 we show the background counts measured all around the studied clusters (solid
dots). We agree fairly well with the measurements by Weir et al. (1995) who also used
DPOSS data but were marginally higher since we include in the background counts also
galaxies belonging to the supercluster in which the studied clusters are embedded in.

Once we have searched for the central $1.5\sigma$ density peak in the central region (cf. Paolillo et al. 2000) we then derive the cluster LF by subtracting from the galaxy counts the local field counts, rescaled to the effective cluster area. This approach allows to take into account the cluster morphology without having to adopt a fixed cluster radius, and thus to apply the local field correction to the region where the signal (cluster) to noise (field and cluster fluctuations) ratio is higher, in order to minimize statistical uncertainties.

\section{Results}

We combine individual LF's of many clusters to obtain a composite  LF for all clusters in
our sample. We adopted the method in  Garilli et al. 1999.  In practice, the composite LF
is obtained by weighting each cluster against the relative number of galaxies in an
opportunely chosen magnitude range (thus taking into account the different degree of
completeness). The final LF for a first sample of 80 clusters is shown in Figure 2.

%
%
%


\begin{figure}[th]
\begin{center}
\epsfysize=12cm 
\epsfxsize=12cm 
\hspace{0.cm}\epsfbox{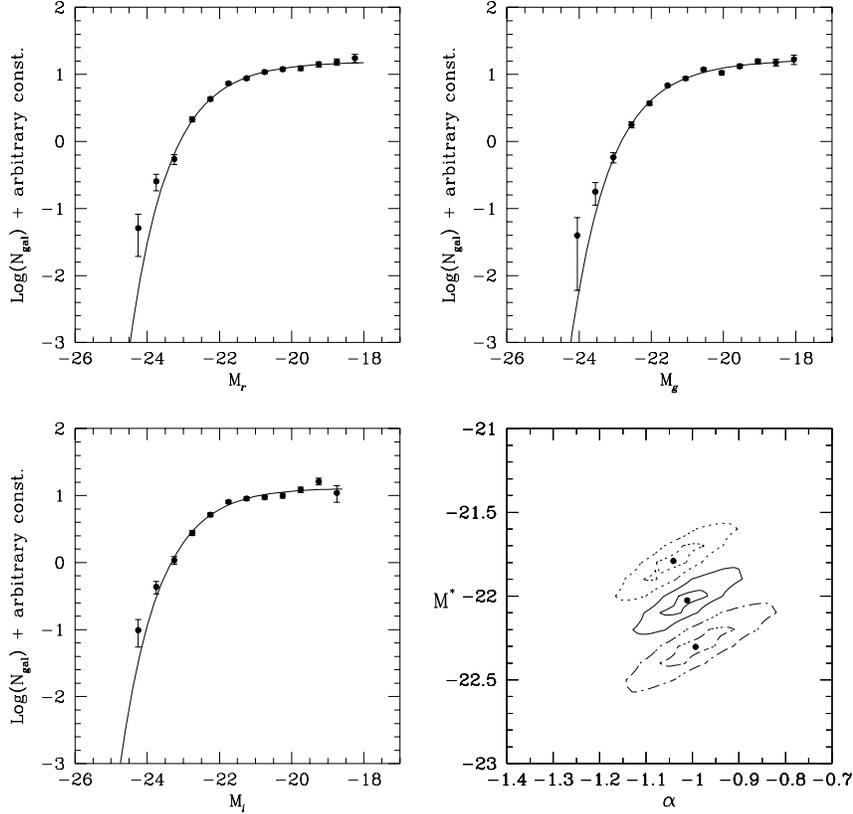} 
\end{center}
\caption[h]{\it{LF for 80 clusters obtained excluding the brightest member of each cluster (filled dots). The best fit Schechter functions are represented by the continuos line, with the 68\% and 99\% confidence levels of the best fit parameters in the bottom right panel ($g$:dotted line; $r$ continuos line; $i$: dashed-dotted line)}.}
\end{figure}

The fit of the LF's to a Schechter (1976) function gives the values: $\alpha =
-1.04^{+0.09}_{-0.07}$; -1.01$^{+0.09}_{-0.07}$; -0.99$^{+0.12}_{-0.11}$ and $M^*=
-21.99^{+0.13}_{-0.17}$; -22.02$\pm$0.16; -22.30$\pm$0.20 mag,  respectively in $g$, $r$
and $i$  where $M^*$ is the characteristic knee magnitude and $\alpha$ is the slope of the
LF at faint magnitudes.  Figure 2 also shows the three best-fit functions together with
the  $68\%$ and $90\%$ confidence levels.

We compared our LF with that found by Garilli (1999) and collaborators and believe that
this comparison is particularly significant since i) they derived their LF from a largely
overlapping set of clusters and in exactly the same photometric system; ii) they used a
completely different criterium for removing interlopers (color changes due to K correction
terms). The comparison is shown in figure 3.

\begin{figure}[t]
\begin{center}
\epsfysize=8cm 
\epsfxsize=8cm 
\hspace{0.cm}\epsfbox{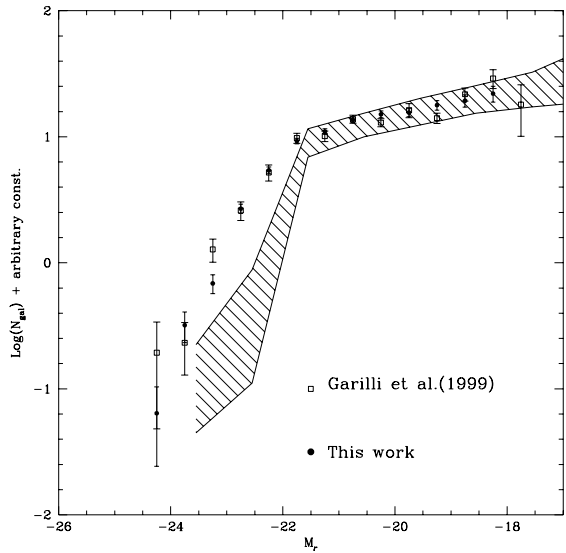} 
\end{center}
\caption[h]{\it{Comparison between our LF, the Garilli et al.(1999) and the Trentham (1998) LF (shaded region) based on CCD data.}}
\end{figure}

In Figure 3 we compare also our LF with that obtained by Trentham (1998). At magnitude
between -22 and -17 the two LF's are in good agreement. We cannot say much about the
sharp rise of the LF found by Trentham at magnitudes fainter than $M_g=-18$, since our
data reach the dwarf range only in the last magnitude bins. On the bright end side of the
LF we find instead a strong difference which may be explained as a result of the fact
that Trentham is understimating the contribution of bright galaxies to the LF.

\begin{figure}[ht]
\begin{center}
\epsfysize=7cm 
\epsfxsize=14cm 
\hspace{0.cm}\epsfbox{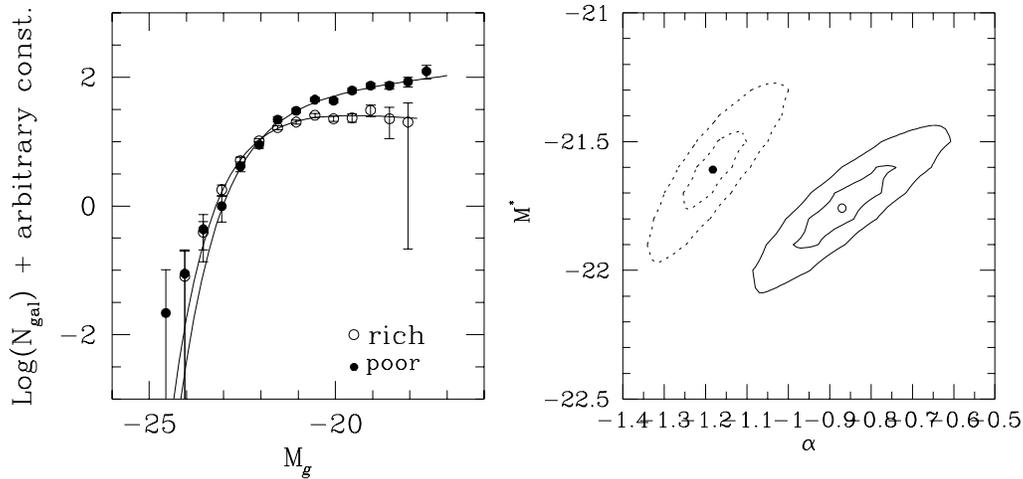} 
\end{center}
\caption[h]{\it{Left panel: the LF and the best fit Schechter functions of the rich 
(R $>$1) and poor ($R \leq 1$) subsamples in the $g$ band. Right panel: 68\% and 99\% confidence levels relative to the
fit parameters ($R > 1$: continous line; $R \leq 1$: dotted line).}}
\end{figure}

The relatively large number of clusters used in the present study allowed us to
investigate the dependence of the LF on the cluster richness parameter. We find (cf.
Figure 4) that with a statistical significance of $3\sigma$, rich (R$>$1) clusters have
shallower faint end than poor (R$\leq1$) ones. This confirm what suggested by Driver,
Couch, \& Phillips (1988) that poor clusters host a larger fraction of dwarfs.

\end {document}